\documentclass{aa}

\usepackage{graphicx}
\usepackage{txfonts}
\usepackage{amsmath}
\usepackage{hyperref}
\usepackage{graphicx}
\hypersetup{
    colorlinks=true,
    linkcolor=blue,
    citecolor=blue,
    urlcolor=blue,
}

\begin{document}

   \title{FarNet-II: An improved solar far-side active region detection method}

   \subtitle{}

   \author{E. G. Broock\inst{\ref{inst1},\ref{inst2}}
          \and
          A. Asensio Ramos\inst{\ref{inst1},\ref{inst2}} 
          \and
          T. Felipe\inst{\ref{inst1},\ref{inst2}}
          }

   \institute{Instituto de Astrof\'{\i}sica de Canarias, 38205, C/ V\'{\i}a L{\'a}ctea, s/n, La Laguna, Tenerife, Spain\label{inst1}
              %\email{}
         \and
             Departamento de Astrof\'{\i}sica, Universidad de La Laguna, 38205, La Laguna, Tenerife, Spain\label{inst2} 
             %\email{}
             }

   \date{Received xxx, aaaa; accepted yyy, aaaa}

% \abstract{}{}{}{}{} 
% 5 {} token are mandatory
 
  \abstract
   {Activity on the far side of the Sun is routinely studied through the analysis of the seismic oscillations detected on the near side using helioseismic techniques such as phase-shift sensitive holography. Detections made through those methods are limited to strong active regions due to the need for a high signal-to-noise ratio. Recently, the neural network FarNet was developed to improve these detections. This network extracts more information from helioseismic far-side maps, enabling the detection of smaller and weaker active regions.}
   {We aim to create a new machine learning tool,  FarNet-II, which further increases FarNet's scope, and to evaluate its performance in comparison to FarNet and the standard helioseismic method for detecting far-side activity.}
   {We developed FarNet-II, a neural network that retains some of the general characteristics of FarNet but improves the detections in general, as well as the temporal coherence among successive predictions. The main novelties of the new neural network are the implementation of attention and convolutional long short-term memory (ConvLSTM) modules. A cross-validation approach, training the network 37 times with a different validation set for each run, was employed to leverage the limited amount of data available. We evaluate the performance of FarNet-II using three years of extreme ultraviolet observations of the far side of the Sun acquired with the Solar Terrestrial Relations Observatory (STEREO) as a proxy of activity. The results from FarNet-II were compared with those obtained from FarNet and the standard helioseismic method using the Dice coefficient as a metric. Given that the application of the ConvLSTM modules can affect the accuracy as a function of the position on the sequence, we take this potential dependency into account in the evaluation.}
   {FarNet-II achieves a Dice coefficient that improves that of FarNet by over 0.2 points for every output position on the sequences from the evaluation dates. Its improvement over FarNet is higher than that of FarNet over the standard method.}
   {The new network is a very promising tool for improving the detection of activity on the far side of the Sun given by pure helioseismic techniques. Space weather forecasts can potentially benefit from the higher sensitivity provided by this novel method.}

   \keywords{Sun:activity - Sun:helioseismology - Sun:oscillations - Sun:sunspots - Sun:UV radiation}

   \maketitle
%
%-------------------------------------------------------------------

\section{Introduction}\label{introduction}

Helioseismology is the branch of heliophysics that uses measurements of the surface oscillations of the Sun to infer the properties of the solar interior. Global helioseismology, the study of the eigenfrequencies of the resonant oscillatory modes, has shed some light on the solar internal structure and its rotation \citep{Christensen-Dalsgaard2002}. The techniques of global helioseismology are, unfortunately, limited to the inference of the global properties of the Sun. As a natural evolution of global helioseismology, local helioseismology \citep{Braun+etal1987,Hill1988,Braun+etal1992,Duvall+etal1993} was developed to infer the properties of local regions of the solar interior or its surface. This is accomplished by studying the whole wave field, instead of just the eigenfrequencies \cite[see][for a review]{Gizon+Birch2005}. 

One of the main techniques under the local helioseismology umbrella is helioseismic holography \citep{Lindsey+Braun1990}. Helioseismic holography applies helioseismic observations of the surface to a solar interior model with no local structure in time reverse, and then it samples the results of the model at various depths. This technique relies on the coherence of the waves on smooth acoustic media to detect structures that disturb their path. Phase-sensitive holography, the technique used in this work, is a specific version of holography that uses the measurement of time-travel perturbations to study active regions that are not directly visible. A more detailed description can be found in \cite{Lindsey+Braun2000}.

Phase-sensitive holography has been applied to the far-side imaging problem, resulting in a technique that is capable of detecting active regions in the non-visible solar hemisphere from the analysis of the near-side wave field \citep{Lindsey+Braun2000b, Braun+Lindsey2001}. Using the seismic data of a region in the visible hemisphere (the so-called “pupil”), the technique infers the wave field in a region of the far side or “focus point” \citep[see][for further details]{Lindsey+Braun2017}. The success of this technique is made possible by the phase shift that active regions introduce between ingoing and outgoing waves from a range of frequencies. This phase shift is fundamentally caused by the Wilson depression \citep{Lindsey+etal2010, Felipe+etall2017}. 
Waves that arrive at regions on the far side where a Wilson depression is present are reflected into the Sun at a deeper layer than those that arrive at the quiet Sun's surface. This shortening of the wave path imprints a negative phase shift upon the arrival of the waves to the near side. The disturbance in the travel time can then be associated with the presence of magnetic activity. We point out that time-distance helioseismology is also being used to perform far-side activity detection  \citep{Duvall+Kosovichev2001,Zhao2007, Ilonidis+etal2009}. 

Although the results are impressive, far-side helioseismic methods are only able to detect strong sunspots due to the low signal-to-noise ratio of the seismic signal from most of the active regions. Smaller and fainter activity is left undetected \citep{GonzalezHernandez+etal2007,Liewer+etal2014, Liewer+etal2017}. 
To bypass the limitations of these techniques, \cite{Felipe+Asensio2019} developed the deep neural network FarNet, a U-net \citep{Ronneberger+etal2015} that improves the sensitivity of phase-sensitive holography, opening the way for the detection of small active regions on the far side. Recently, \cite{Broock+etal2021} confirmed FarNet's reliability and its superior performance to that of the standard method. The goal of this paper is to present FarNet-II, a new architecture that further improves FarNet predictions by implementing convolutional long short-term memory (LSTM) modules and attention mechanisms into a U-net architecture. 

The LSTM architectures \citep{LSTM} are recursive neural networks that were developed as an improvement over plain recurrent neural networks \citep[RNNs;][]{Rumelhart1986LearningIR}. Recurrent neural networks are architectures used to compute predictions over time series of data. They are especially useful for cases where temporal coherence is important (i.e., for forecasting problems). These types of networks take inputs recursively and use weights extracted from operations with previous inputs to compute the next outputs. Since RNNs are especially susceptible to the vanishing gradient problem, particularly when used over long time series \citep{HS}, LSTMs were proposed.

The second main novelty of FarNet-II is the use of attention mechanisms. They are a variety of algorithms that focus on weighting the importance of different parts of the data running through a neural network, and on using their relative importance to optimize the performance of the network.
These tools started by being applied to translation neural networks \citep{bahdanau2016neural, vaswani2017attention}, but their range of applicability has since become much wider. One of the fields in which these mechanisms have been used is computer vision, improving models for image classification \citep{hu2019local, hu2019squeezeandexcitation} and semantic segmentation \citep{fu2019dual, li2019expectationmaximization}. 

The paper is organized as follows: Sect. \ref{Det} explains methods of detecting far-side activity, including FarNet-II; Sect. \ref{R} presents the obtained results; and those results are discussed in Sect. \ref{DC}, along with the conclusions of the paper.

\section{Detection of far-side activity}\label{Det}

\subsection{Phase-sensitive seismic method}

The detection of activity on the far side of the Sun is founded on the analysis of far-side phase-shift maps. Helioseismic holography is employed to compute these maps from continuous near-side Doppler data. The latter are acquired from synoptic observations \citep[the Global Oscillation Network Group, GONG,][]{Harvey+etal1996} or space-based observatories \citep[Helioseismic and Magnetic Imager, HMI][]{Schou+etal2012}. In this work, we focus on HMI far-side seismic maps, which are regularly published by the Joint Science Operation Center (JSOC)\footnote{\url{http://jsoc.stanford.edu/ajax/lookdata.html}}. Two different data products with a cadence of 12 hours are available: one that uses 24 hours of Doppler data for its generation and another one that uses five days of Doppler data instead. 

The detection of active regions on the far side is routinely carried out by Stanford's Strong Active Region Discriminator (SARD)\footnote{\url{http://jsoc.stanford.edu/data/far-side/explanation.pdf}}. This process uses the phase-shift maps computed with five days of Doppler data from HMI, and searches for those regions with a phase-shift lower than $-0.085$ rad. Then, the integral over the region's area, in millionths of a hemisphere ($\mu$Hem), is calculated. This magnitude is the so-called seismic strength ($S$). The presence of a seismically detected far-side active region is claimed for regions with $S>400$ $\mu$Hem\,rad \citep{Liewer+etal2017}. 

\subsection{FarNet}

FarNet was the first successful attempt at using deep learning to improve the interpretation of far-side phase-shift maps for activity detection. Activity detection can be seen as a binary semantic segmentation task, in which every pixel of the output needs to be classified as active or non-active. It is usually achieved by using deep convolutional neural networks (DCNNs). One of the architectures that is more commonly employed to pursue this task is U-net. This architecture is based on an encoder-decoder structure that uses tools such as convolutional layers, batch normalization \citep[BN;][]{Ioffe+etal2015}, and rectified linear unit activation functions \citep[ReLU;][]{Nair+Hinton2010}. The encoder reduces the spatial size of the images in successive steps via max-pooling \citep{Goodfellow-et-al-2016}, while the number of channels is increased. This is done to extract and combine spatial information at many scales. 
On the decoder, the opposite happens, and the channel dimension is reduced while the spatial size is gradually increased to the spatial size of the input, via interpolation. U-net uses skip connections between the encoder and the decoder to better utilize multiscale information, and improve performance during training and evaluation.

\begin{figure*}[!tbp]
    \centering
  \includegraphics[width=8.3cm]{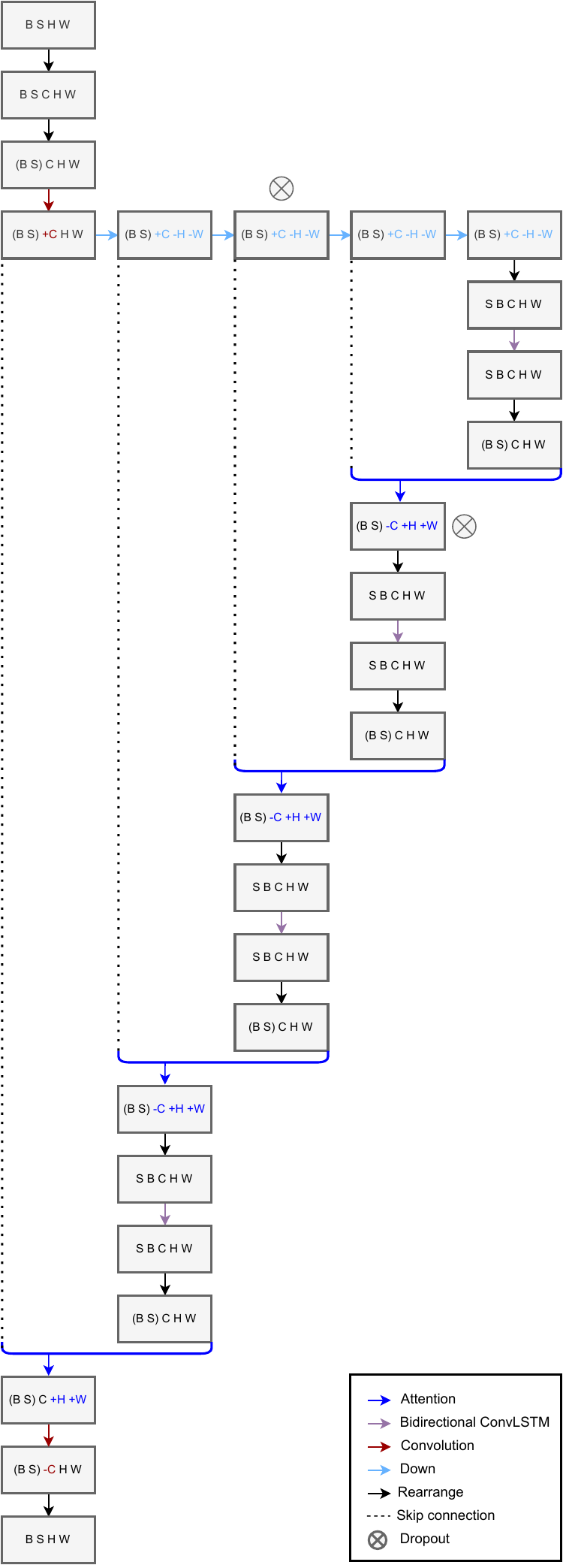}
    \centering
  \caption{\textbf{General representation of FarNet-II.} Original dimensions are batch (B), sequence (S), height (H), and width (W). Channel dimension (C) is additionally used in the following steps. Batch and sequence dimensions are joined together for every operation in the network, except for the application of the ConvLSTM modules. The red arrows symbolize simple convolutions, the light blue arrows symbolize the downward operation that reduces the spacial size of the images while increasing the number of channels, the violet arrows symbolize the bidirectional convolutional LSTM modules, and the dark blue arrows symbolize attention mechanisms. $\otimes$ symbolizes a dropout of 0.5.}
  \label{NN}
\end{figure*}

FarNet takes, as input, sequences of 11
phase-shift maps with a temporal cadence of 12 hours, each of them computed in a 24-hour window of Doppler data (in contrast to SARD detections, which are currently based on phase-shift maps obtained from five days of Doppler data). A region spanning 120$^\circ$ in longitude and 144$^\circ$ in latitude is included in each map. The output is a probability map of the same far-side region from the central date of the input. Its values are constrained to the $[0,1]$ range using a sigmoid function at the output of U-net.
As proposed by \cite{Felipe+Asensio2019}, and later verified by \cite{Broock+etal2021}, it is
important to post-process the output of FarNet to infer which features of the output are reliable active region detections. First, a Gaussian filter is applied over the outputs, with a full width at half maximum of 1.5 pixels. Then, regions with five contiguous pixels with a probability value higher than 0.2 are selected. Finally, for each detected region, the integrated probability ($P_{i}$) is calculated. This quantity is the integral of the probability over the regions' area, measured in deg$^2$. A region with an integrated probability of $P_{i}>100$ is taken as a reliable detection. 

\begin{figure*}[!tbp]
    \centering
  \includegraphics[width=9cm]{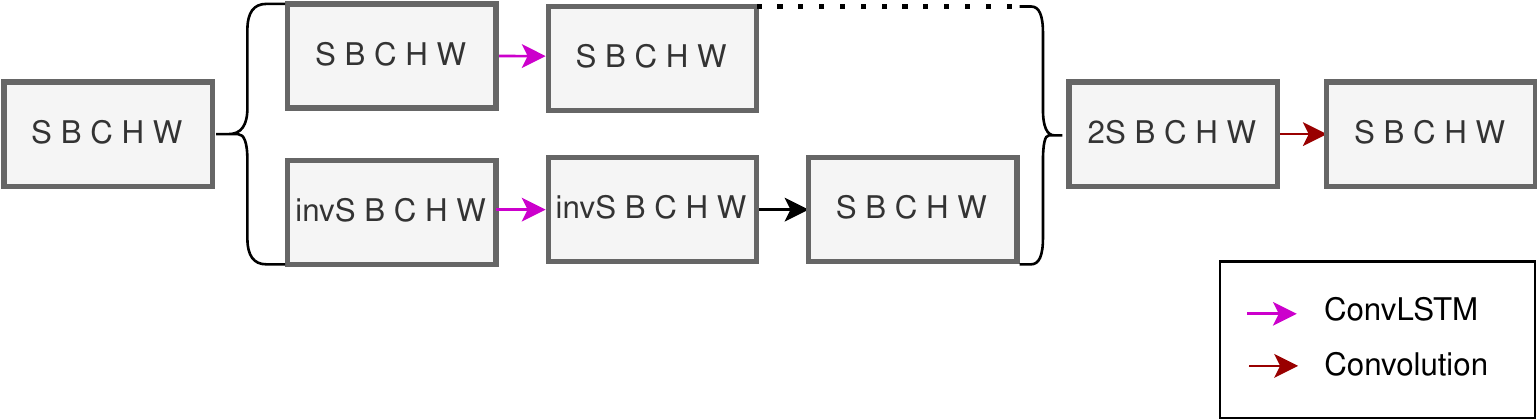}
    \centering
  \caption{\textbf{Bidirectional ConvLSTM module.} Before the application of this module, the batch and sequence dimensions are split, and the input is duplicated but inverted in the sequence dimension. A ConvLSTM module is applied over both tensors, and the result of the application over the inverted one is re-inverted and concatenated with the result in the forward direction. The output of the module is obtained with a convolution over the concatenated tensors, which gives an output with the same dimension as the input.}
  \label{fig:convlstm}
\end{figure*}

\subsection{FarNet-II}
In this paper, we develop FarNet-II, an evolution of FarNet that greatly enhances its capabilities. It maintains some structural properties of FarNet, but it introduces some improvements, such as bidirectional convolutional LSTM modules, attention mechanisms, and dropout. One of the most relevant improvements is that FarNet-II can now produce one activity prediction per input date, instead of one only prediction for the central date of the input. This is achieved through the application of bidirectional ConvLSTM modules on specific parts of the network, which exploit the time coherence both forward and backward in time. 

Figure \ref{NN} shows a graphical representation of our new model. It resembles a standard
U-net architecture, but with the addition of the attention and ConvLSTM blocks in the decoder. We use, as input, sequences
of 11 phase-shift maps of spatial size $H \times W$. The encoder is applied in parallel to all the input
maps of the sequences. For this reason, we combine the batch and sequence dimensions to fully exploit the parallelization capabilities of the hardware we use for training. The encoder increases the number of channels per map from one to $C$. Spatial dimensions are reduced by the consecutive application of a max-pooling and two convolutional layers with a ReLU activation function. Once the encoder operations are fully applied, the output of the encoder is passed through the decoder. Every operation, except those of the ConvLSTM layers, is applied in parallel to all elements in the input sequences. The decoder upsamples the information in the lowest spatial scale and combines it with the information obtained from the encoder at the same spatial resolution thanks to an attention layer. This process goes on until the decoder is applied over every scale. Figure \ref{fig:convlstm} shows how the decoder uses the ConvLSTM layer, which keeps memory from the previous elements of the sequence during prediction, in a bidirectional way, keeping also memory from the next elements. We explain the details of these two innovations in the following sections. Dropout is also applied to two locations in the network, one on the encoder and the other on the decoder. The output of the network goes through a sigmoid to limit its values to the [0,1] range before exiting the U-net architecture. The described model and an example of usage can be found online\footnote{GitHub repository:  \url{https://github.com/EBroock/FarNet-II}}.

\subsubsection{LSTM and convolutional LSTM}

As a recursive network, the main innovation of the LSTM \citep{LSTM} over the RNN is the presence of a memory cell, $c_{t}$, which modulates the information from previous inputs used to compute the next output. The memory cell is modified thanks to three self-parameterized gates: the input, the forget, and the output gates. Information of the input currently being processed by the layer is included in the cell if the input gate is open. Likewise, the values from the previous cell state are forgotten according to the value of the forget gate. The output gate controls if the latest cell state is propagated onto the next hidden state. The status of all gates (open or closed) is determined by sigmoid activation functions. 

The LSTM modules have been used for a variety of purposes, including computer vision, with high success, but as their algorithm was not purposefully developed to tackle that task, their initial implementations did not take into account the possible existence of spatial correlations between different positions of the image. Long short-term memory was initially applied to computer vision in a fully connected manner, having to unfold the inputs into one-dimensional vectors, losing important spatial information. Architectures as convolutional layers naturally take into account these relations, especially when used over various resolutions of the inputs, as in encoder-decoder architectures.

\cite{shi2015convolutional} developed an encoder-decoder architecture for the task of predicting the future rainfall intensity in a local region over a short time period (precipitation nowcasting), in which convolutions are directly introduced as an intrinsic part of LSTM layers. Since its conception, ConvLSTM modules have been used in machine learning architectures for multiple applications, such as the detection of violence in videos \citep{HansonViolence} or predictions of urban expansion \citep{boulila2021novel}. 

For our work, we use a bidirectional version of the ConvLSTM proposed by \cite{shi2015convolutional}, which consists of two ConvLSTM layers. As shown in Fig. \ref{fig:convlstm}, the first layer is applied to the sequence in the original order, while the second is applied to the reversed sequence. After reversing again the result of the application of this second layer, both sequences are concatenated and a convolutional layer reduces the output to the input size.

\subsubsection{Attention}

Attention mechanisms on FarNet-II are used to modulate the skip connections before using them at each step of the decoder. The module uses the tensor on the decoder previous to its application and the corresponding skip connection. Both tensors are processed through convolutions, ReLU activations, and a final sigmoid, which produces a mask. This mask is then applied to the original skip connection, which is used on the decoder in the traditional manner of U-net. For this work, we used the implementation given by \citet{Fillioux2020} of the attention module proposed by \citet{oktay2018attention}. 

\begin{figure*}
    \centering
  \includegraphics[width=16cm]{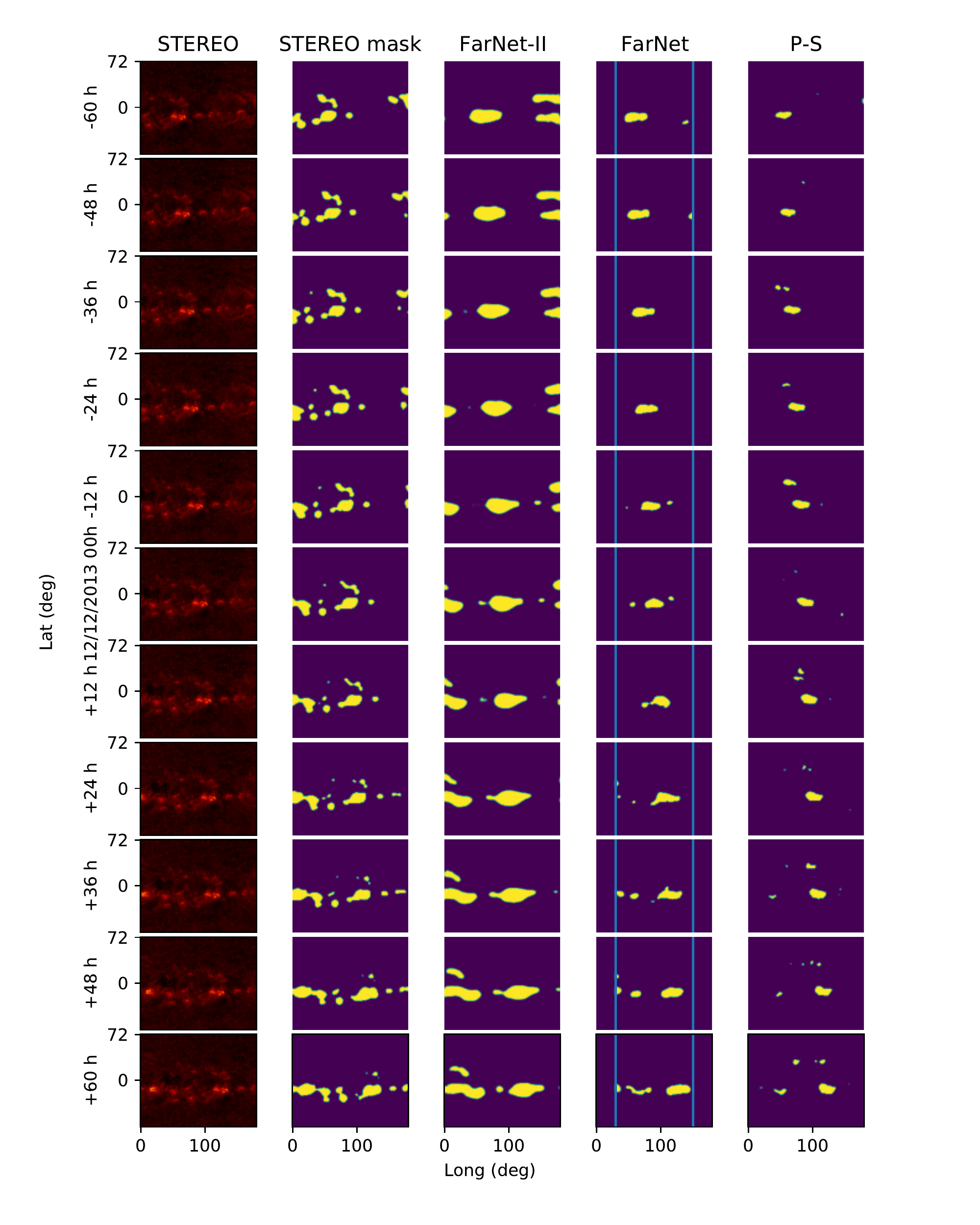}
  \caption{Comparison of an output sequence, centered on December 12, 2013, for each method. The first column shows the square root of the STEREO data used to compute the activity masks. The second column shows the activity masks. Third to fifth columns show outputs from FarNet-II, FarNet, and the phase-sensitive method, respectively, for the region corresponding to the EUV masks on the second column. In columns 2 to 5, the color gradient goes from purple for zero or values near zero to yellow for values near one or one. Outputs from FarNet are only valid on the central range of 120 degrees of longitude (vertical blue lines). Seismic strength and integrated probability were not taken into account to select the regions on the outputs from FarNet and the phase-sensitive method. Every region that passed the post-processing was included.}
  \label{comp_filt}
\end{figure*}

\subsubsection{Dropout}
Dropout is a regularization method developed by \cite{dropout_sri}, whose goal is to improve the generalization of neural network capabilities, reducing overfitting to a specific training set. This is achieved by randomly ignoring some nodes on certain layers of the model during the training process. For FarNet-II, we applied dropout to 50\% of the nodes in two different stages of the data flow. 
\subsubsection{Post-processing}

As in FarNet, filtering was applied to the outputs of FarNet-II to get rid of small disturbances in the background that do not correspond to active regions. We applied Gaussian filtering with a full width at half maximum of 1.5 pixels, and every pixel in regions with five contiguous pixels with a probability higher than 0.2 was set to one, while the rest of the pixels were set to zero. This made the comparison with STEREO masks more rigorous. The difference in the training metrics before and after applying the filtering is negligible.

\subsubsection{Extreme Ultraviolet data}
FarNet-II goes through supervised training by forcing the output to be as close to the desired
target as possible. These targets are obtained from a binarization process of 304 {\AA} Carrington maps from the Solar Museum Server of NASA \citep{Liewer+etal2017}. These maps join together images from the Extreme Ultraviolet Imager \citep[EUVI;][]{Wulser+etal2004} on board the Solar Terrestrial Relations Observatory \citep[STEREO;][]{Kaiser2004}, and from the Atmospheric Imaging Assembly \citep[AIA;][]{Lemen+etal2012} on board the Solar Dynamics Observatory \citep[SDO;][]{Pesnell+etal2012}. The training only employs the region of the maps corresponding to the far-side hemisphere, that is, the data acquired with STEREO. Precedents of extreme ultraviolet (EUV) image usage as a proxy of far-side activity can be found in \citet{Liewer+etal2012}, \citet{Liewer+etal2014}, and \citet{Zhao+etal2019}.
The process through which the EUV maps were binarized is explained in detail by \citet{Broock+etal2021}.

\subsubsection{Training}\label{train}

Dates from December 4, 2011 to August 18, 2014, were included in the training. The inputs were sequences of 11 consecutive far-side phase-shift maps, from which a region of 144$^\circ$ in latitude and 180$^\circ$ in longitude was taken, centered on the far side. As a target, we used the corresponding 11 binarized EUV masks of the same region. 
The total number of available input-target pairs with good far-side coverage was 2253. We note that the size of this training set is very limited. We partially solved this issue by using augmenting. It consisted of vertically flipping the images and the expected values. Due to the lack of a larger dataset, we carried out the study using a cross-validation method, dividing the training set into segments of 60 elements and choosing one of those elements as validation on each run. In total, 37 independent trainings were performed, each of them with 4340 input-target pairs for training and 120 input-target pairs for validation  (including augmenting). We made evaluations on the validation set of each training, and then we averaged the resulting metrics among them. 

The training of the model was done by minimizing the Dice loss computed between the outputs and the EUV binary masks. The Dice loss derived from the Dice coefficient. The Dice coefficient is a measure of the superposition of arrays of data with binary labels. For this specific scenario, it accounts for the accuracy of pixel labeling and can take values from 0 (no overlapping between pixel values on the output and on the associated EUV activity mask) to 1 (complete overlapping of output and the EUV activity mask). Since the values of both the outputs and the EUV masks are restricted to lying between 0 and 1, the Dice coefficient is given by:
\begin{equation}\label{eq:1}
D = \frac{2\sum_{b,x,y} i(x,y,b) o(x,y,b)+\epsilon}{\sum_{x,y,b} i(x,y,b)+\sum_{x,y,b} o(x,y,b)+\epsilon},
\end{equation}
where $i(x,y,b)$ and $o(x,y,b)$ are the values of the output and target images for all pixels ($x$ and $y$ coordinates) and elements of the minibatch ($b$ coordinate), while $\epsilon$ is a small quantity (0.001 in our case) that prevents the Dice coefficient from being undefined. To use the Dice coefficient as an efficient loss function for the training of a model, the simplest way is subtracting it from the unit, computing the Dice loss, given by:
\begin{equation}\label{eq:2}
L_{D} = 1-D.
\end{equation}
We implemented the model and trained it using the open-source PyTorch library \citep{Paszke+etal2019},
and optimized the Dice loss using the Adam optimizer \citep{kingma2017adam}  with a learning rate of 3$\times$10$^{-4}$ during ten epochs and a batch size of 10. These ten epochs were sufficient for the Dice loss to stop increasing on validation data.

\begin{figure*}%[!tbp]
    \centering
  \includegraphics[width=16cm]{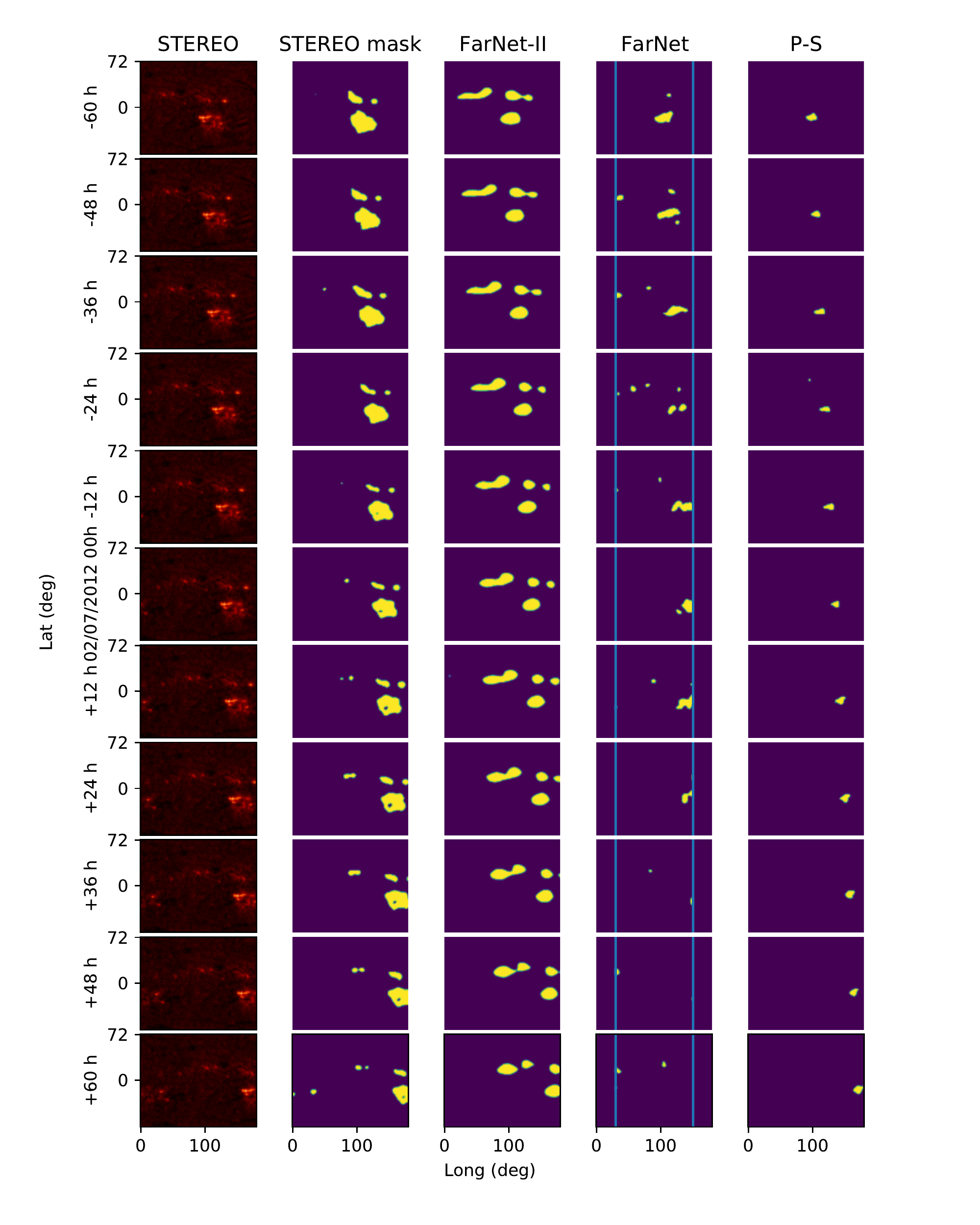}
  \caption{Same as for Fig. \ref{comp_filt}, but for data centered on July 2, 2012.}
  \label{comp_filt_2}
\end{figure*}

\section{Results}\label{R}

\begin{figure*}[!t]
   \centering
   \begin{minipage}{0.48\textwidth}
     \includegraphics[width=1\linewidth]{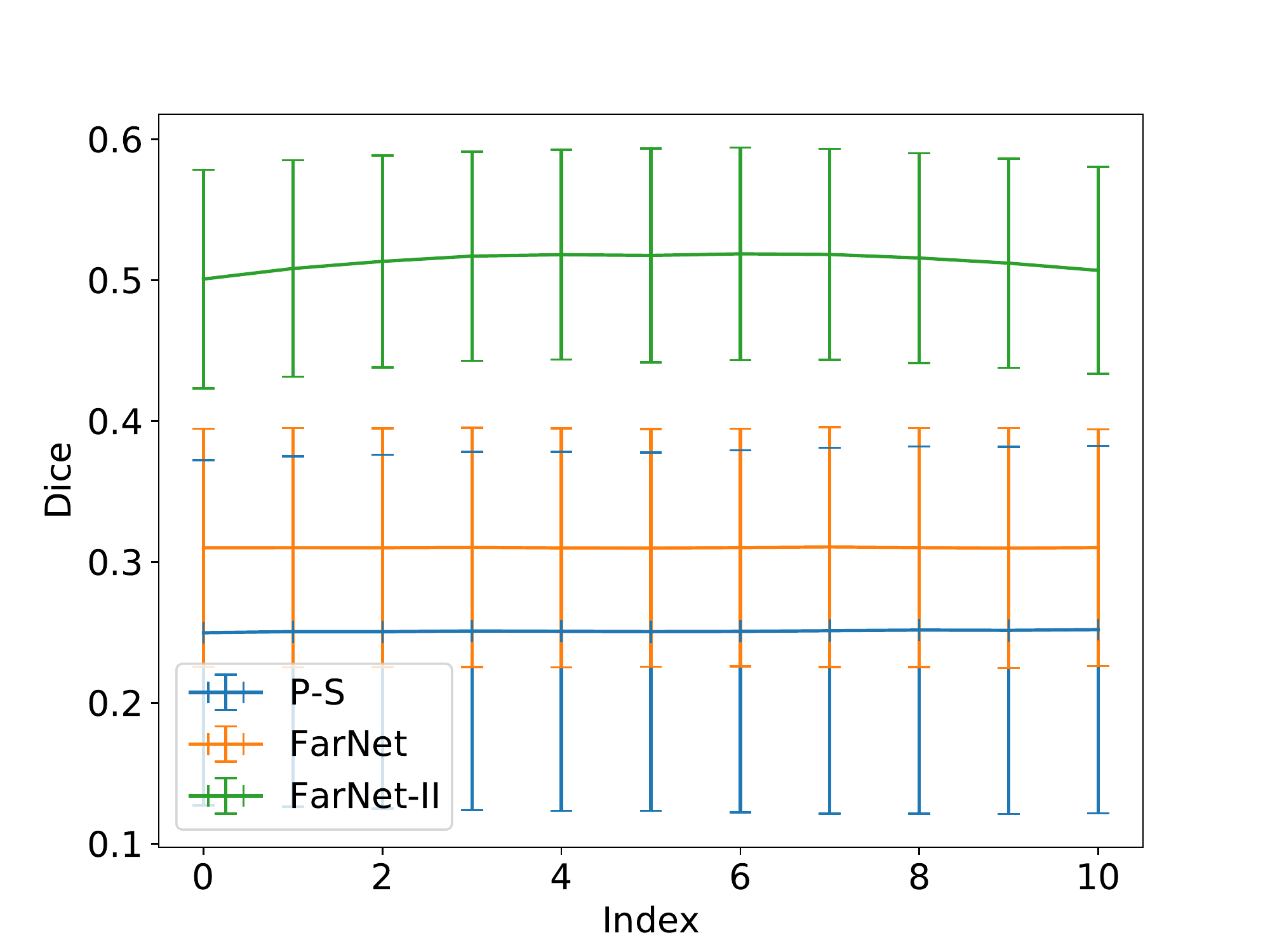}
     \caption{Dice coefficient per method, as a function of the position of the images on the output sequences of 11 elements. Thresholds for reliable detection for FarNet ($P_{i} > 100$) and the phase-sensitive method ($S>400$) were not taken into account. For FarNet, every region on outputs from FarNet with more than five contiguous pixels and a probability over 0.2 was used to compute the value. The vertical bars represent the standard deviation of the mean Dice coefficient over the 37 validation sets used in the study.}\label{Fig:dice_completo_1}
   \end{minipage}\hfill
   \begin{minipage}{0.48\textwidth}
     \includegraphics[width=1\linewidth]{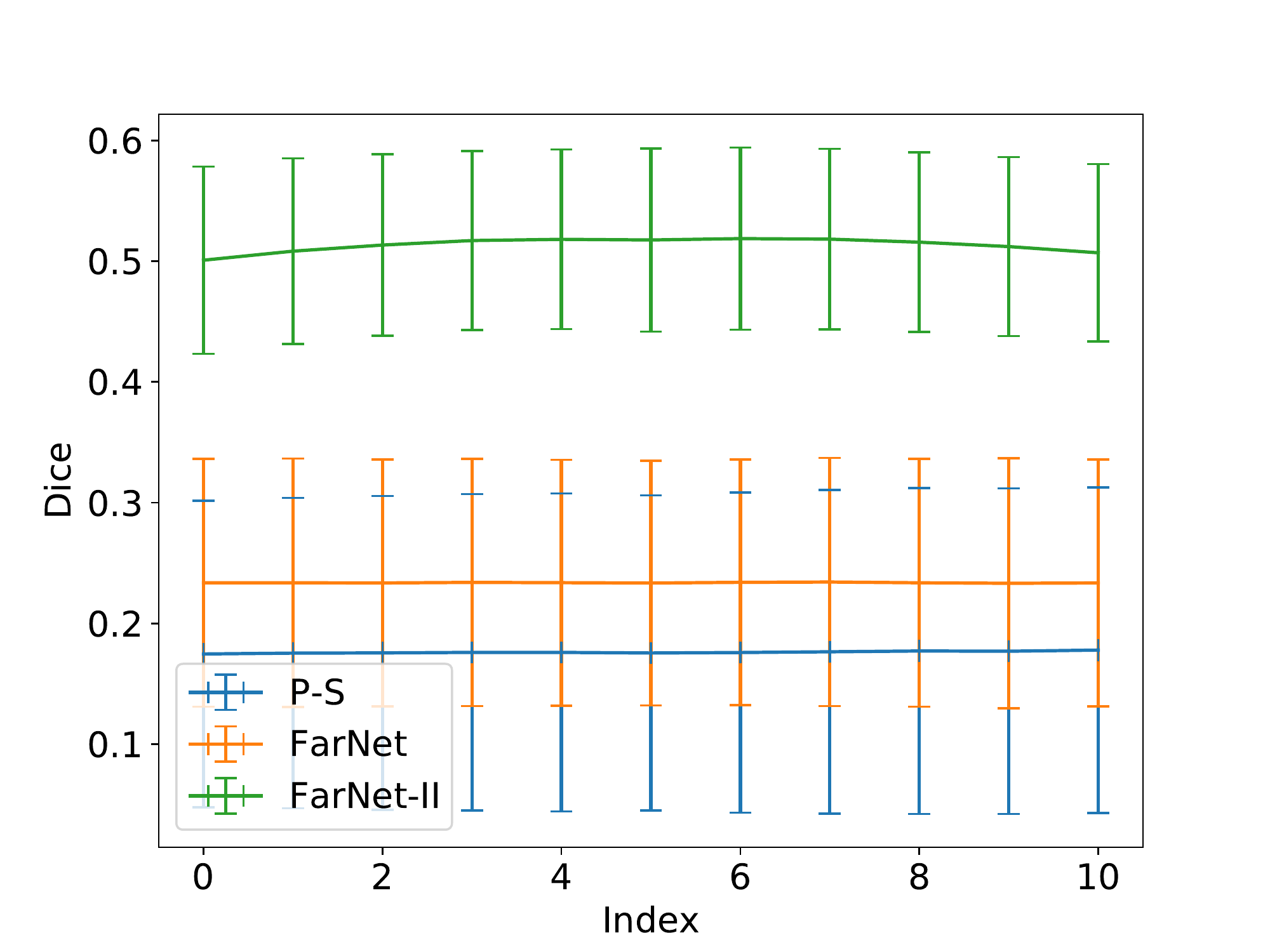}
     \caption{Dice coefficient per method, as a function of the position of the images on the output sequences of 11 elements. Only regions with $P_{i}$>100, for FarNet, and with $S$>400, for the phase-sensitive method, are taken into account. The vertical bars represent the standard deviation of the mean Dice coefficient over the 37 validation sets used in the study.}\label{Fig:dice_completo_2}
   \end{minipage}
   \centering
\end{figure*}

To characterize the reliability of all the methods (standard phase-sensitive holography, FarNet, and FarNet-II) and compare their performance, we analyzed the values of the Dice coefficient between the EUV binary masks and the outputs of each method for each validation set. While FarNet-II produces a prediction spanning 180$^\circ$ in longitude, FarNet infers a shorter range. For this reason, the comparison was only made on the far-side section common to every method, spanning 120$^\circ$. 

In our previous paper \citep{Broock+etal2021}, we used a different metric, in which each blob detected on the output of the model was assumed to be a different object. By comparison with the STEREO activity masks, we distinguished whether it was a true or a false detection. However, FarNet-II outputs tend to be less segmented, with various blobs merged, which makes the previous method of comparison less optimal.

\subsection{Qualitative comparisons}
We start by qualitatively comparing the outputs of all methods for two sequences, shown in Figs. \ref{comp_filt} and 
\ref{comp_filt_2}. Each output batch of ten sequences takes between 13 and 14 seconds to be produced by the trained network on a CPU. The first columns show the square root of the STEREO EUV maps, with the second column displaying the STEREO masks defined using our procedure. The third, fourth, and fifth columns show the results obtained by FarNet-II, FarNet, and the standard phase-sensitive helioseismic method, respectively. Each row shows data of an element of the sequence, representing an instant in the temporal evolution of the far side for the dates of study. The solar rotation can be seen in the masks and also in the predictions. It is clear from these figures that the predictions of FarNet-II are much more accurate than those of the other methods when compared with the STEREO activity masks. FarNet-II can correctly predict small activity regions while also detecting many of the large regions. We speculate that the prediction of these regions is possible thanks to the time coherence exploited by the ConvLSTM in FarNet-II.
The results of FarNet-II displayed in Figs. \ref{comp_filt} and \ref{comp_filt_2} show an excellent prediction ability, which is in strong contrast with the poorer results from FarNet and the classical phase-sensitive helioseismic method.

\subsection{Quantitative comparisons}
The visual inspection of FarNet-II predictions demonstrates a very good prediction ability. In this section, we employ quantitative comparison methods to show that this is true for the large majority of cases in the validation sets.

This quantitative comparison is made using the Dice coefficient as a metric, which is then averaged over the validation sets of every training. We analyzed how the metric behaves for all of the considered methods. For an in-depth comparison, we checked the metrics when the thresholds in $P_{i}$ and $S$ (seismic strength) proposed for a reliable detection were considered or not. 
The global results are shown in Table \ref{table:1}. The global prediction capabilities of all methods were measured with the Dice coefficient averaged over every element of the sequences and all validation sets. FarNet-II clearly has improved performance, with the Dice coefficient increasing by more than 0.2 points over the most performant of all previous methods.

\begingroup
\renewcommand*{\arraystretch}{1.2}
\begin{table}[!h]
\caption{Average of the Dice coefficients over every sequence element, for each method and model, including variations in the filtering on outputs from FarNet ($P_{i}$ value) and the phase-sensitive method ($S$ value). The standard deviation of the results for every validation set is shown in the third column.}
\label{table:1}
\centering 
\begin{tabular}{l r r }
\hline
 Method & Dice & Std\\
\hline
   P-S S>0 & 0.251 & 0.13\\
   P-S S>400 & 0.176 & 0.13\\
   FarNet P$_{\textrm{\textit{i}}}$>0 & 0.310 & 0.08\\
   FarNet P$_{\textrm{\textit{i}}}$>100 & 0.234 & 0.10\\
   FarNet-II & \textbf{0.513} & \textbf{0.08}\\

\hline
\end{tabular}
\end{table}
\endgroup

Since FarNet-II produces a prediction for all times using the information of previous and later input images of the sequence, we expect the reliability of the prediction to be potentially different for different times inside the sequence. We analyzed this in detail by calculating the Dice coefficient for each element of the sequence on the validation set. Since FarNet and the standard phase-sensitive method do not make use of this temporal information, the results of each index were computed with the appropriate window of phase-shift maps centered on the specific date of each element of the sequence. The results are shown in Fig. \ref{Fig:dice_completo_1} when no special threshold is used. When the optimal thresholds are considered ($P_{i}>100$ for FarNet and $S>400$ for the phase-sensitive method), the results are those shown in Fig. \ref{Fig:dice_completo_2}. It is important to note that very restrictive values were selected for these optimal thresholds \citep{Broock+etal2021}. They guarantee the presence of an EUV emission counterpart to the seismic active region with a 96\% confidence level. When they are taken into account, the Dice coefficient exhibits poorer values since actual detections are discarded due to their lower confidence. According to these results, FarNet-II produces a Dice coefficient that is almost a factor of two larger than those of the other methods, on average. Additionally, this consistency is maintained for all indices of the sequence, with only a small improvement in the predictions for the central frames.

\subsection{Ablation study}
\begin{figure*}[!t]
   \centering
   \begin{minipage}{0.48\textwidth}
     \includegraphics[width=1\linewidth]{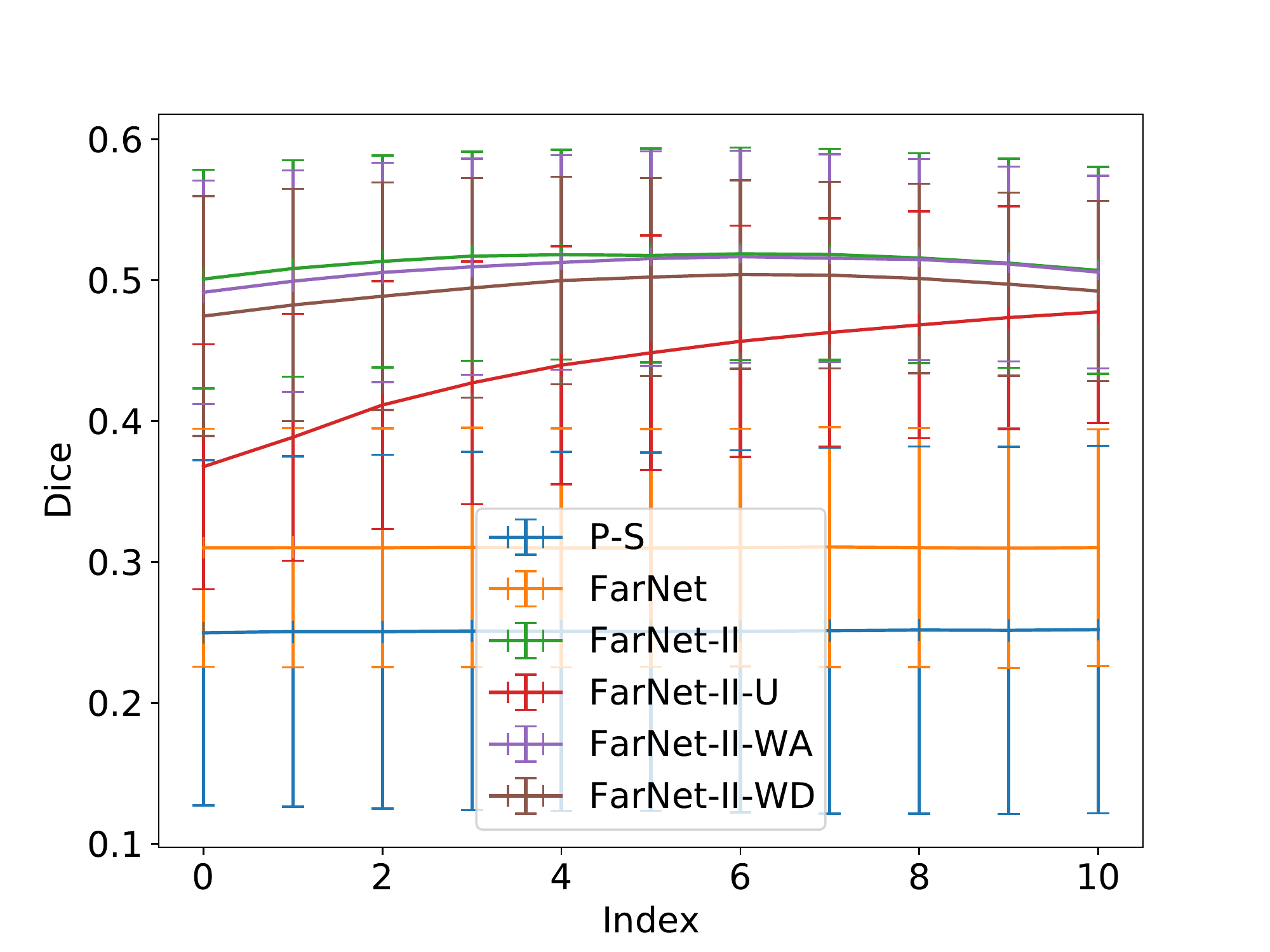}
     \caption{Dice coefficient per method, as a function of the position of the images on the output sequences of 11 elements. This figure includes the ablated versions of FarNet-II. Thresholds for reliable detection for FarNet ($P_{i} > 100$) and the phase-sensitive method ($S>400$) were not taken into account. For FarNet, every region on outputs from FarNet with more than five contiguous pixels and a probability over 0.2 was used to compute the value. The vertical bars represent the standard deviation of the mean Dice coefficient over the 37 validation sets used in the study.}\label{Fig:dice_unidir_1}
   \end{minipage}\hfill
   \begin{minipage}{0.48\textwidth}
     \includegraphics[width=1\linewidth]{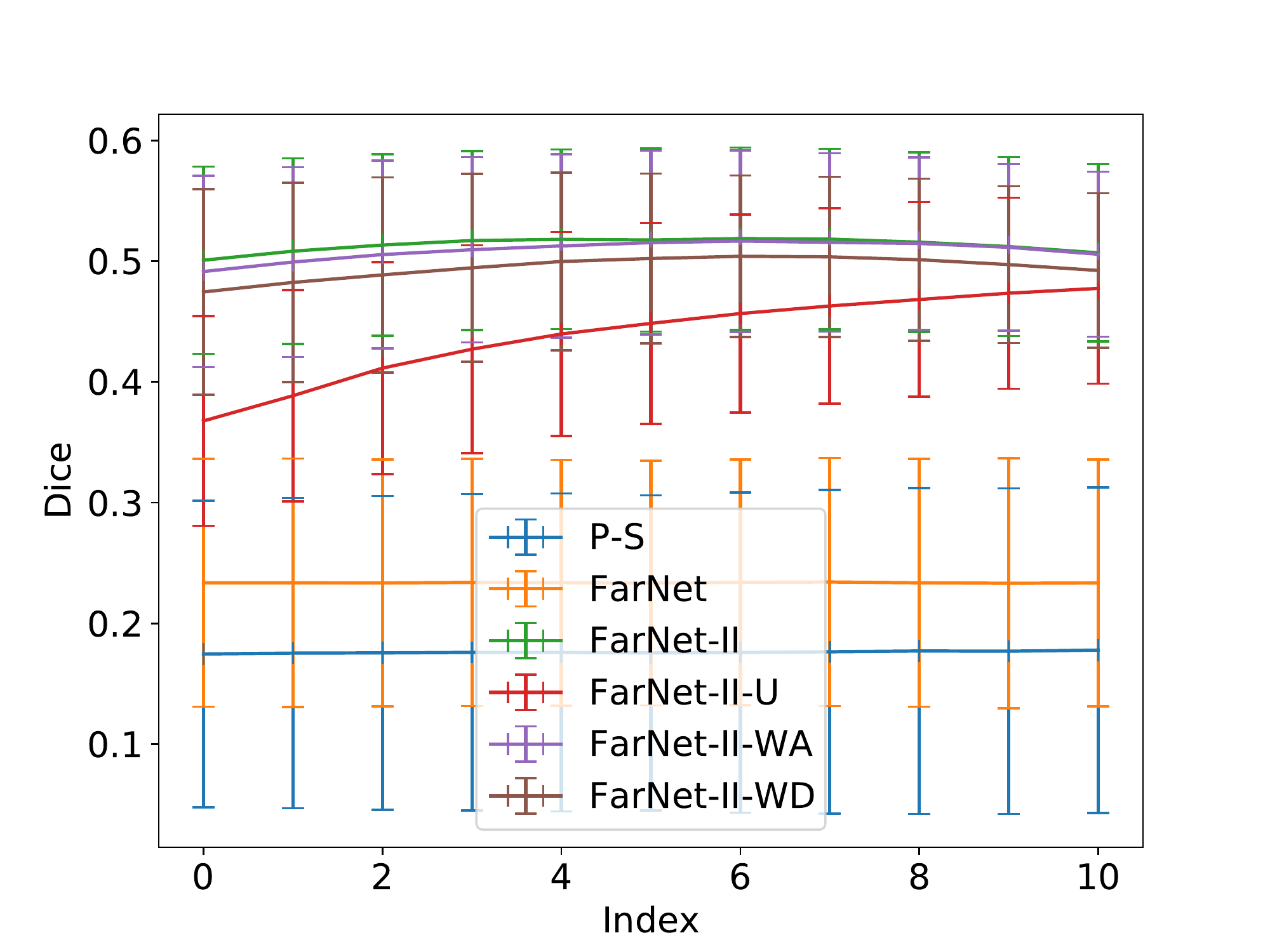}
     \caption{Dice coefficient per method, as a function of the position of the images on the output sequences of 11 elements. This figure includes the ablated versions of FarNet-II. Only regions with $P_{i}$>100, for FarNet, and with $S$>400, for the phase-sensitive method, are taken into account. The vertical bars represent the standard deviation of the mean Dice coefficient over the 37 validation sets used in the study.}\label{Fig:dice_unidir_2}
   \end{minipage}
   \centering
\end{figure*}

We carried out an ablation study to determine the relative importance of each of the new layers added to FarNet-II concerning the previous FarNet neural network. This study, although limited, demonstrates that the final architecture arguably produces the best predictions. To this end, we trained different models: 1) using unidirectional ConvLSTM layers instead of the bidirectional ones employed in the final model; 2) removing the attention layers; and 3) removing the regularizing effect of dropout. The average Dice  coefficients computed on the validation sets for each model are represented in Figs. \ref{Fig:dice_unidir_1} and \ref{Fig:dice_unidir_2}. The production time of outputs does not vary significantly for ablated versions of FarNet-II. The total averaged Dice coefficient achieved by each model can be found in Table \ref{table:2}. The results demonstrate that adding recursion produces the largest 
improvement over the baselines, with attention and dropout increasing the prediction ability only marginally, although still monotonically. In the case of using the unidirectional ConvLSTM, we see that the prediction for the elements at the end of the sequence is better than those at the beginning, demonstrating that exploiting the time correlation in the two directions is important.

\begingroup
\renewcommand*{\arraystretch}{1.2}
\begin{table}%[h]
\caption{Average of the Dice coefficients of every sequence element for FarNet-II and its ablated models (\textbf{U}: unidirectional ConvLSTM; \textbf{WA}: without attention; \textbf{WD}: without dropout). The standard deviation of the results for every validation set is shown in the third column.}
\label{table:2}
\centering 
\begin{tabular}{l r r}
\hline
 Method & Dice & Std\\
\hline
   FarNet-II & \textbf{0.513} & \textbf{0.08}\\
   FarNet-II-U & 0.438 & 0.08\\
   FarNet-II-WA & 0.509 & 0.07\\
   FarNet-II-WD & 0.495 & 0.07\\

\hline
\end{tabular}
\end{table}
\endgroup

\subsection{Performance as a function of longitude}

Figure \ref{dice_long} illustrates the dependence of the performance of FarNet-II on longitude. This study was individually performed for every index in the sequence of outputs. Additionally, the total average, including all the outputs from the sequence, is shown for completeness.

The results exhibit a marked variation with the sequence index. At the beginning of the series (indices 0-2), far-side regions with a lower longitude (that is, the solar region that has just rotated onto the far-side hemisphere) show a higher Dice coefficient. In contrast, a low Dice coefficient is retrieved at high longitudes (regions that are about to rotate onto the visible hemisphere). A progressive change in this trend is found as higher indices are considered. The variation in the Dice coefficient with the longitude exhibits a mostly flat profile at the middle of the series (indices 3-5), whereas in the last steps of the sequence (indices 6-10), the Dice coefficient increases with the longitude.

This dependence on the longitude is consistent with the displacement of the active regions across the far-side hemisphere due to the solar rotation. Active regions located at high longitudes in the first elements of the series quickly rotate onto the visible hemisphere, and their presence is no longer tracked by seismic far-side maps. Similarly, active regions with a low longitude in the last steps of the series have just rotated onto the far side and only appear in those last steps. In summary, a weaker performance is found in those two extremes (high longitude and low index, low longitude and high index) due to the limited information available in the input sequence of far-side seismic maps. These results also support the relevance of exploiting the bidirectional temporal correlation.

\section{Discussion and conclusions}\label{DC}

We have presented a new neural architecture, FarNet-II, which combines some characteristics of the original FarNet \citep{Felipe+Asensio2019} with the use of bidirectional Conv-LSTM, attention modules, and dropout. We have proven that this model further improves the capabilities of FarNet in the detection of far-side activity. 

FarNet-II was trained using activity masks extracted from EUV data from the far side as expected values. This is an improvement over FarNet's training, where we used near-side binarized magnetograms from half a rotation later than the inputs to the network. Even though these magnetograms were processed to eliminate every active region emerging on the near side, they do not provide an accurate characterization of the far-side activity at the temporal period when the seismic maps were computed, since the size and shape of the regions can vary between both dates, and regions may have decayed before rotating onto the near side. In these new training sets, far-side activity fed to the network is obtained from direct far-side observations, and it is strictly co-temporal to the seismic maps, leading to higher accuracy. 

The enhanced performance of FarNet-II, as compared with other methods, has been proven through the visual inspection of their outputs in comparison with the actual far-side activity captured by EUV observations from STEREO (Figs. \ref{comp_filt} and \ref{comp_filt_2}). We have also performed a quantitative comparison by employing the Dice coefficient as a metric to evaluate the similarity between the predicted far-side regions and the actual activity in EUV maps. The results show a remarkably higher value of the Dice coefficient for FarNet-II, strongly supporting its improved performance.

\begin{figure*}[!t]
    \centering
  \includegraphics[width=16cm]{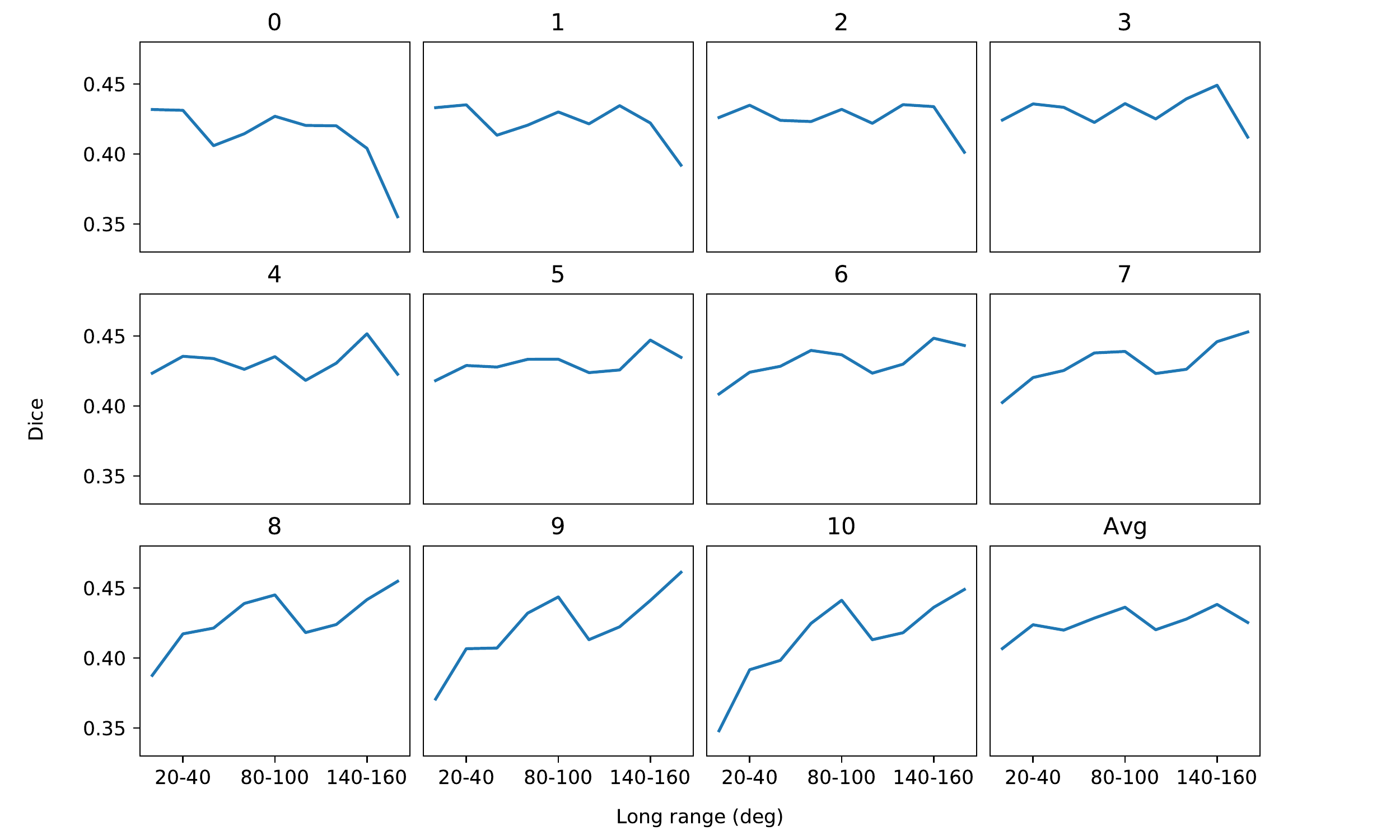}
  \caption{Average over every validation set of the Dice coefficient on FarNet-II's outputs as a function of the longitude range in degrees. Each range covers 20º. Each panel represents the Dice coefficient on the sequence element indicated above it. The bottom right panel represents the average of all sequence elements. A longitude of $0^\circ$ corresponds to the west limb, $90^\circ$ to the center of the far-side hemisphere, and $180^\circ$ to the east limb.}
  \label{dice_long}
\end{figure*}

We have evaluated the individual contribution from each of the new ingredients implemented in FarNet-II. The analysis clearly points to the ConvLSTM modules as the main driver of the increased reliability of the new architecture. Interestingly, the version of FarNet-II with only unidirectional ConvLSTM modules (forward in time) exhibits an upward trend of the Dice coefficient as the sequence index increases (Fig. \ref{Fig:dice_unidir_1}). This is due to the increase in information that later elements on the sequence receive from previous iterations. When bidirectional ConvLSTM is implemented (full FarNet-II), the performance is similarly good for the whole sequence, except for longitudes near the limb at the beginning and the end of the sequence (Fig. \ref{dice_long}). The other novel modules implemented in FarNet-II (attention modules and dropout) provide a modest improvement in its performance. All in all, the best results are found when all these new layers act together. Although not shown explicitly in this work, we checked the coherence of the differential rotation for the sequence of FarNet-II's outputs. The results are consistent with the measured solar differential rotation.

We consider this work as a new step forward to improve the imaging of the far side of the Sun. Nowadays, direct far-side images can only be acquired by Solar Orbiter \citep{SolarOrbiter} during some periods of its orbit, even though they are fundamental for space weather applications (the branch of astrophysics dedicated to studying the Sun and the state of the interplanetary space on the Solar System). The global photospheric magnetic field, not only in the visible hemisphere, is necessary in order to model the heliospheric magnetic field and solar wind. Currently, synoptic maps that complete the magnetism in the non-visible hemisphere with near-side observations obtained many days in advance are generally employed. This approximation of the far-side magnetism can be updated with sophisticated flux transport models \citep{Schrijver+DeRosa2003}. However, these models cannot account for new active regions emerging on the far side or active regions that keep growing after rotating onto the non-visible hemisphere. The detection of activity on the far side is important to completely characterize the global photospheric magnetic field and, thus, the heliosphere. \citet{Arge+etal2013} proved that incorporating seismically detected far-side active regions into the modeling of the heliosphere produces results that better match in situ measurements of the solar wind. Our study provides a remarkable improvement of the capabilities of local helioseismology to characterize the far-side magnetism, getting us closer to the goal of implementing these techniques in space weather forecasting applications. The main limitation of this study is the lack of data to construct bigger training sets. We have been forced to work with small validation sets that can lead to somehow noisy statistics. We plan to bypass this limitation with new data in future works.

\begin{acknowledgements}
We thank P. C. Liewer and collaborators for making publicly available the composite STEREO/EUVI and SDO/AIA maps necessary to carry out this research. We also thank C. Cid, from Universidad de Alcalá de Henares (UAH), for her invaluable ideas for future projects using FarNet-II. Financial support from grants PGC2018-097611-A-I00 and PID2021-127487NB-I00, funded by MCIN/AEI/ 10.13039/501100011033 and by “ERDF A way of making Europe”, and grant PROID2020010059 funded by Consejería de Economía, Conocimiento y Empleo del Gobierno de Canarias and the European Regional Development Fund (ERDF) is gratefully acknowledged. TF acknowledges grant RYC2020-030307-I funded by MCIN/AEI/ 10.13039/501100011033 and by “ESF Investing in your future”. We acknowledge the community effort devoted to the development of the following 
open-source packages that were
used in this work: \texttt{numpy} \citep[\texttt{numpy.org},][]{numpy20}, 
\texttt{matplotlib} \citep[\texttt{matplotlib.org},][]{matplotlib}, \texttt{PyTorch} 
\citep[\texttt{pytorch.org},][]{Paszke+etal2019}, and \texttt{SunPy} 
\citep[\texttt{sunpy.org},][]{sunpy_community2020},
\texttt{einops} \citep{rogozhnikov2022einops}, and \texttt{h5py} \citep{hdf5}.
\end{acknowledgements}

 \bibliographystyle{aa} 
 \bibliography{biblio.bib}

\end{document}